\documentclass[namedreferences]{SolarPhysics}
\usepackage[optionalrh]{spr-sola-addons} 

\usepackage{graphicx,url,color} 

\usepackage[pdfborder={0 0 0 },urlcolor=blue,breaklinks]{hyperref}
\ifx \doiurl  \undefined \def \doiurl#1{\href{http://dx.doi.org/#1}{\url{#1}}}\fi
\ifx \adsurl  \undefined \def \adsurl#1{\href{http://adsabs.harvard.edu/abs/#1}{\url{#1}}}\fi

\newcommand{\etal}{{\it et al.}}
\newcommand{\eg}{{\it e.g.}}
\newcommand{\degr}{\mbox{$^\circ$}}%
\newcommand{\kw}{\mbox{$k$}--\mbox{$\omega$}\ }%

\newcommand{\apj}{    {\it Astrophys. J.}}

\newcommand{\nat}{    {\it Nature}}

\newcommand{\pasj}{   {\it Pub. Astron. Soc. Japan}}
\newcommand{\solphys}{{\it Solar Phys.}}
\newcommand{\ssr}{    {\it Space Sci. Rev.}}

\begin{document}
\begin{article}
\begin{opening}

\title{Analysis of Helioseismic Power-Spectrum Diagram of A Sunspot} 
\author{Junwei \surname{Zhao}$^{1}$ \sep
       Dean-Yi \surname{Chou}$^{2}$ }
\runningauthor{J.~Zhao, D.-Y.~Chou}
\runningtitle{Sunspot \kw Diagram}
\institute{$^{1}$ W.~W.~Hansen Experimental Physics Laboratory, 
          Stanford University, Stanford, CA 94305-4085, USA
          email: \href{mailto: junwei@sun.stanford.edu}
                              {junwei@sun.stanford.edu} \\
          $^{2}$ Physics Department, National Tsing Hua University, 
          Hsinchu, Taiwan email: \href{mailto: chou@phys.nthu.edu.tw} 
                                              {chou@phys.nthu.edu.tw}
         }

\begin{abstract}
The continuous high spatial-resolution Doppler observation of the Sun 
by {\it Solar Dynamics Observatory / Helioseismic and Magnetic Imager} 
allows us to compute helioseismic \kw power-spectrum diagram using only 
oscillations inside a sunspot. Individual modal ridges can be clearly 
seen with reduced power in the \kw diagram constructed by use of 
40-hour observation of a stable and round sunspot. Comparing with the \kw 
diagram obtained from a quiet-Sun region, inside the sunspot the 
$f$-mode ridge gets more power reduction than $p$-mode ridges, especially
at high wavenumber. The $p$-mode ridges all 
shift toward lower-wavenumber (or higher-frequency) areas for a given 
frequency (or wavenumber), implying an increase of phase velocity beneath 
the sunspot. This probably results from acoustic waves' travel 
across the inclined magnetic field of the sunspot penumbra. Line-profile 
asymmetries exhibited in the $p$-mode ridges are more significant in 
the sunspot than in quiet Sun. Convection inside the sunspot is also 
highly suppressed, and its characteristic spatial scale is substantially 
larger than the typical convection scale of quiet Sun. These observational 
facts demand a better understanding of magnetoconvection and interactions 
of helioseismic waves with magnetic field.
\end{abstract}


\end{opening}

\section{Introduction}
Analysis of helioseismic power-spectrum diagrams, also known as \kw 
diagrams, obtained from solar oscillation signals, is a useful way to 
derive solar atmospheric properties, interior structures, and subsurface 
dynamics (\eg\ \opencite{rho97}; \opencite{sch98}). While many global-scale 
properties of the Sun were inferred by analyzing \kw diagrams obtained 
from the global-scale observations, some local-scale properties were 
derived by analyzing \kw diagrams obtained from oscillations in 
local areas, known as ring-diagram analysis \cite{hil88}. Solar areas
containing active regions have been studied using ring-diagram techniques
to infer interior structures \cite{bas04, bal11} and subsurface
flow fields \cite{hab04, kom05} of active regions. However, the \kw
diagrams used in these analyses are largely composed of oscillations
outside of sunspots rather than those inside sunspots, hence
represent properties of large areas instead of properties of confined 
sunspot areas.  Other local helioseismology techniques, \eg, time-distance
helioseismology \cite{duv93b, kos00, giz09}, also need to apply filters 
over \kw diagrams although not to analyze the \kw diagrams directly.

Despite the scientific interest and importance of constructing \kw diagrams 
using only oscillations inside sunspots, this has not been successfully
achieved due to either poor spatial resolution or limited temporal coverage. 
\inlinecite{pen93} attempted to compute \kw diagrams of two sunspots using 
observations from Mees Solar Observatory, but were not able to obtain distinct
modal ridges. Using {\it Hinode} Ca~H observations, \inlinecite{nag07} studied
acoustic power maps inside a sunspot, but did not analyze its \kw 
diagram due to the short temporal coverage. Employing Hankel's decomposition, 
\inlinecite{bra87} analyzed the \kw diagrams made from acoustic waves 
traveling into and out from sunspots separately, but these diagrams 
were not computed using oscillations observed directly inside sunspots. 
SOHO/MDI \cite{sch95} full-disk data did not have sufficient spatial
resolution to construct \kw\ diagrams using only signals inside sunspots, 
while its high-resolution data lacked sufficient coverage time. 
The {\it Helioseismic and Magnetic Imager} \cite{scherrer12, sch12} 
onboard {\it Solar Dynamics Observatory} (SDO/HMI) provides 
continuous observations of 
the Sun with a high spatial resolution and steady temporal cadence, 
and this allows us to perform a \kw diagram analysis for sunspots. 
In this article, we report some interesting
properties found in the \kw diagram of a sunspot. We introduce our 
data-analysis procedure in Section 2, and present our analysis results in 
Section 3. We discuss these results in Section 4, and give conclusions in 
Section 5.

\section{Data Analysis}
\label{sec2}

\begin{figure}
\centerline{\includegraphics[width=0.98\textwidth]{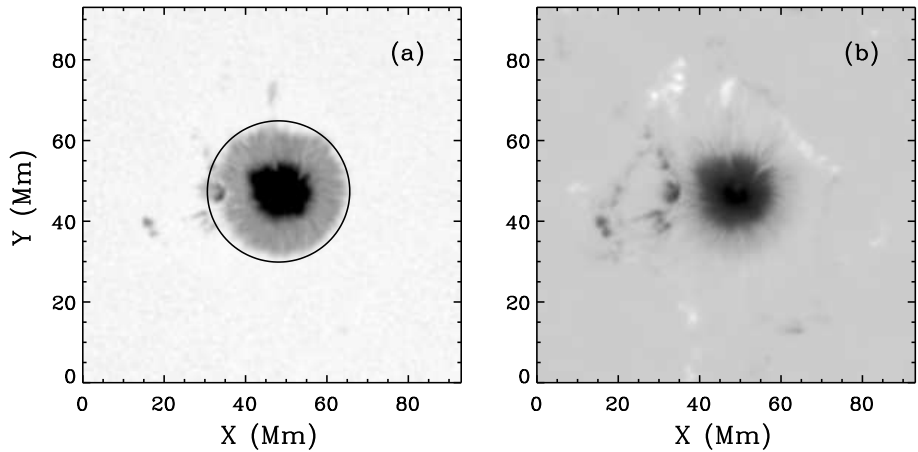} }
\caption{Continuum intensity image (a) and line-of-sight component of 
the magnetic field (b) for AR~11092. The magnetic-field strength is 
displayed with a range of $-2000$ to 500 Gauss. Oscillations 
observed inside the black circle in panel (a) are used to calculate
the \kw diagram for the sunspot. }
\label{spot}
\end{figure}

SDO/HMI has continuous observations of the full solar disk except 
during some short eclipse periods, with a temporal cadence of 45 seconds and 
a spatial resolution of 1.0 arcsec. Dopplergrams observed by HMI, 
primarily used for helioseismology studies as recently demonstrated 
by \inlinecite{zha12}, are used in this study. The sunspot located 
inside NOAA AR~11092 is selected for the \kw power-spectrum analysis. 
This sunspot, as displayed in Figure~\ref{spot}, was located at 
the latitude of $12.4\degr$N, and remained stable and round in shape 
during the period of analysis. Magnetic field surrounding this sunspot 
was relatively simple. A period of 40 hours, covering 16:00 UT 2 August  
through 08:00UT 4 August, 2010, is selected for analysis, and this is 
roughly corresponding to the sunspot's central meridian crossing, 
from $8.8\degr$ East of the central meridian to $8.8\degr$ West of it. 
This 40-hour period is then divided into five eight-hour segments, and each 
data segment is analyzed separately. The results obtained from each 
segment are then averaged to give the final results. Every eight-hour data 
segment is tracked with the sunspot's rotation speed as given 
by Howard (1990), and is remapped 
using Postel's projection with the remapping center at the sunspot center
and with a sampling resolution of $0.03\degr$ pixel$^{-1}$, slightly 
oversampling in most areas except near the solar-disk center. 

Fourier transform is used to convert data from space -- time 
domain to Fourier domain, and \kw diagram is constructed in the 
three-dimensional Fourier domain. To avoid possible complications 
of Fourier analysis due to the irregular shape of the sunspot boundary, 
we choose to use the oscillations inside a circular boundary 
as denoted by the black circle in Figure~\ref{spot}a. When performing 
Fourier transforms, we keep the entire box size unchanged as shown in 
Figure~\ref{spot} and fill up the area outside of the circle with zeroes,
and this is a normal practice known as zero-padding in Fourier
analysis. The area near the boundary of the inside and outside of 
the circle is tapered with a cosine bell to avoid sharp transition that
may give spurious signals. For the \kw diagram made following the above 
procedure, the power at low-wavenumber areas gets contaminated due to 
the small box size and the circular structure with data surrounded by zeroes. 
Thus, we do not take signals lower than $k = 0.3$ rad~Mm$^{-1}$ into 
the analysis that follows.

To better understand the analysis results, it is useful to do similar 
analyses over quiet-Sun regions for comparison. Here, we select a piece of
quiet Sun that has a same disk location and a same temporal duration 
as the analyzed sunspot covering the period of 16:00 UT 17 August 
through 08:00UT 19 August, 2010, fifteen days after the sunspot 
analysis period. The exactly same analysis procedures are applied on 
these data segments, and results are used as references to compare 
with the results from the sunspot.

\section{Results}
\label{sec3}

\begin{figure}[!t]
\centerline{\includegraphics[width=0.7\textwidth]{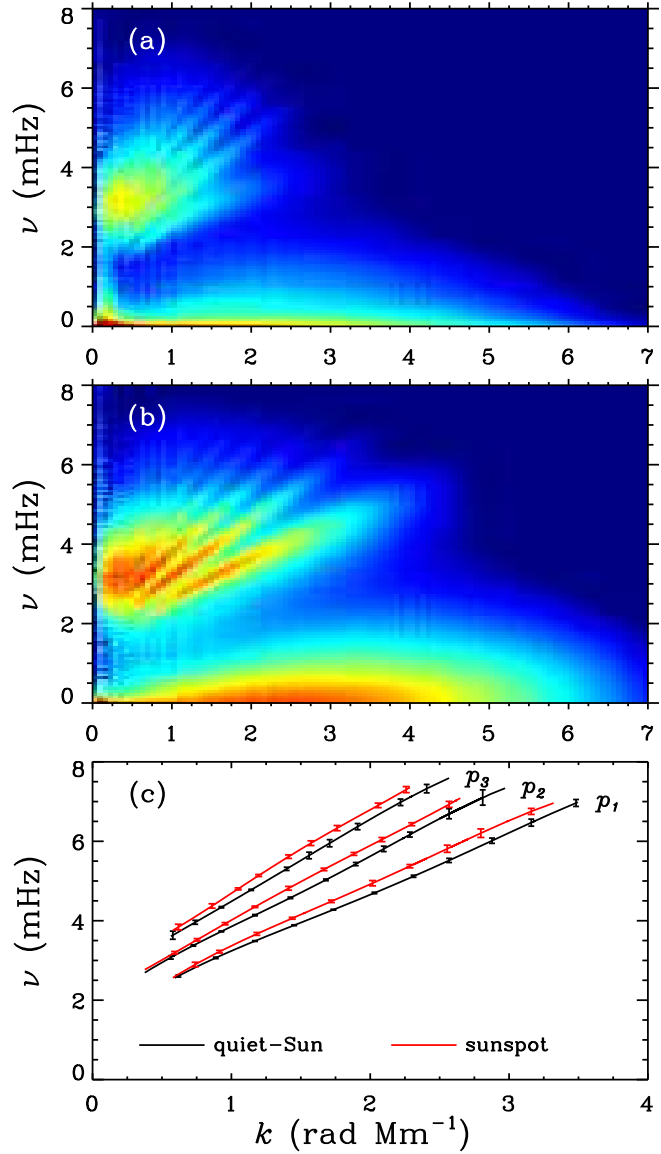} }
\caption{Panels (a) and (b) display the \kw diagrams obtained from
the sunspot and from the quiet-Sun region, respectively. Color scales in
these two panels are the same. Panel (c) shows the ridge-peak locations,
obtained by fitting both \kw diagrams, to better illustrate the relative 
locations of the modal ridges.}
\label{kw}
\end{figure}

\begin{figure}
\centerline{\includegraphics[width=1.0\textwidth]{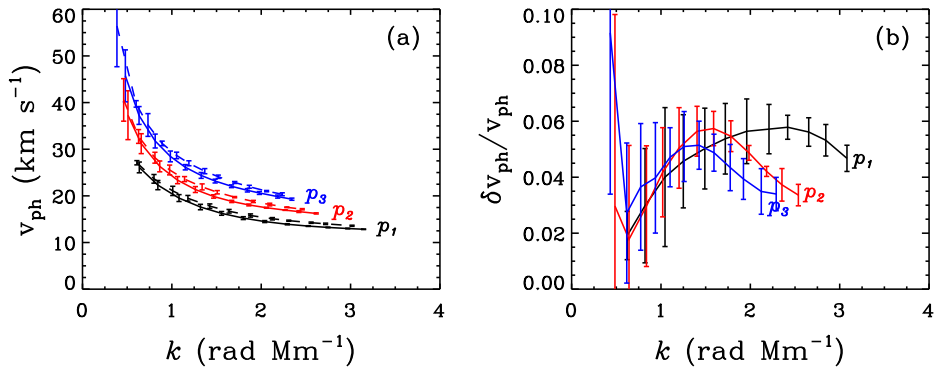} }
\caption{(a) Phase velocities calculated from fitted power ridges 
$p_1 - p_3$, with solid lines for the quiet-Sun region and dashed lines
for the sunspot. (b) Perturbations of phase velocity obtained from 
the sunspot relative to that obtained from the quiet-Sun region. }
\label{phv}
\end{figure}

Figure~\ref{kw}a and b display the \kw diagrams calculated from the sunspot 
region and the quiet-Sun region, respectively. Individual modal ridges can 
be clearly seen in both diagrams. As expected, the helioseismic power of 
each ridge is much weaker in the diagram obtained from the sunspot than 
that from the quiet-Sun. Compared with the $p$-mode (acoustic mode) power 
reductions for the sunspot, the power reduction in the $f$-mode 
(surface-gravity mode) ridge is more substantial, and the length of this ridge 
seems more shortened due to the higher power reduction at
higher wavenumber. For both \kw diagrams, we fit the $p_1 - p_3$ 
modal ridges by peak-finding to better illustrate how the location of each 
individual ridge obtained from the sunspot differs from that obtained 
from the quiet-Sun region. The fitting results are shown in Figure~\ref{kw}c. 
It is difficult to get a reliable fitting for the $f$-mode ridge of 
the sunspot. The three fitted modal ridges obtained from the sunspot 
all shift to the lower wavenumber (or alternatively speaking, 
higher frequency) side relative to the ridges obtained from the quiet Sun 
for a given frequency (or wavenumber). The shift amount is not uniform 
along each individual ridge, and the largest shift is up to 0.3~mHz. 

\begin{figure}
\centerline{\includegraphics[width=0.95\textwidth]{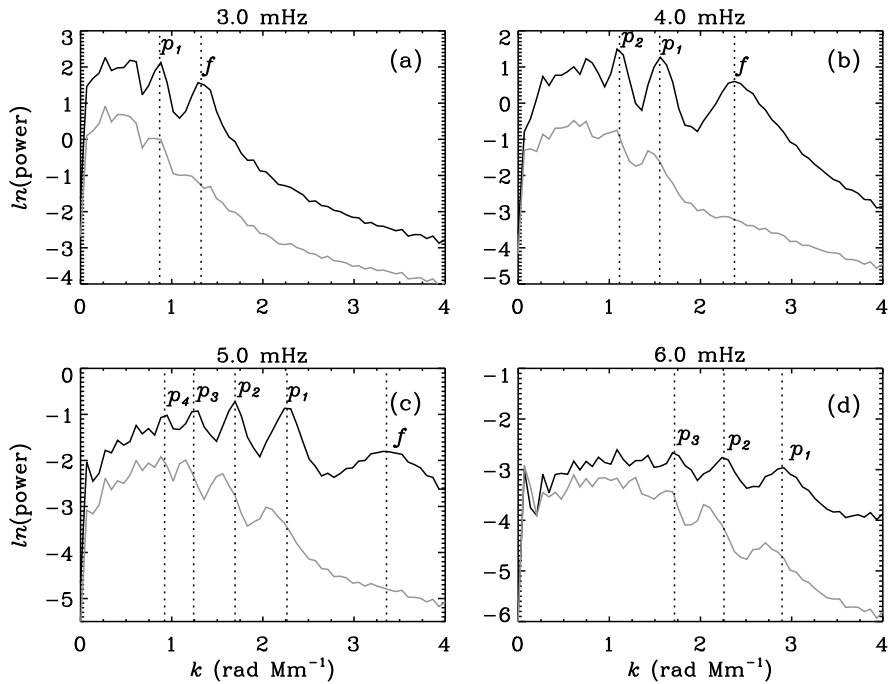} }
\caption{Comparisons of power spectrum at different frequencies obtained 
from the sunspot (gray lines) and the quiet-Sun region (black lines). 
The vertical dashed lines indicate the locations of the ridge peaks for 
the quiet Sun. } 
\label{kw_cut}
\end{figure}

Based on the fitted modal ridges shown in Figure~\ref{kw}c, we calculate 
phase velocity $v_{\mathrm{ph}} = \omega /k$ along each ridge of both 
diagrams and display the results in Figure~\ref{phv}a. The derived phase 
velocity in the sunspot region is often larger than in the quiet-Sun 
region by up to 3 km~s$^{-1}$. Figure~\ref{phv}b shows 
the phase-velocity perturbation for the sunspot relative to the quiet-Sun
region $\delta v_{\mathrm{ph}} / v_{\mathrm{ph}}$ for each modal 
ridge. The calculated $\delta v_{\mathrm{ph}} / v_{\mathrm{ph}}$, 
although large in error bars, mostly fall in range of 0.02 to 0.06.

\begin{figure}
\centerline{\includegraphics[width=0.95\textwidth]{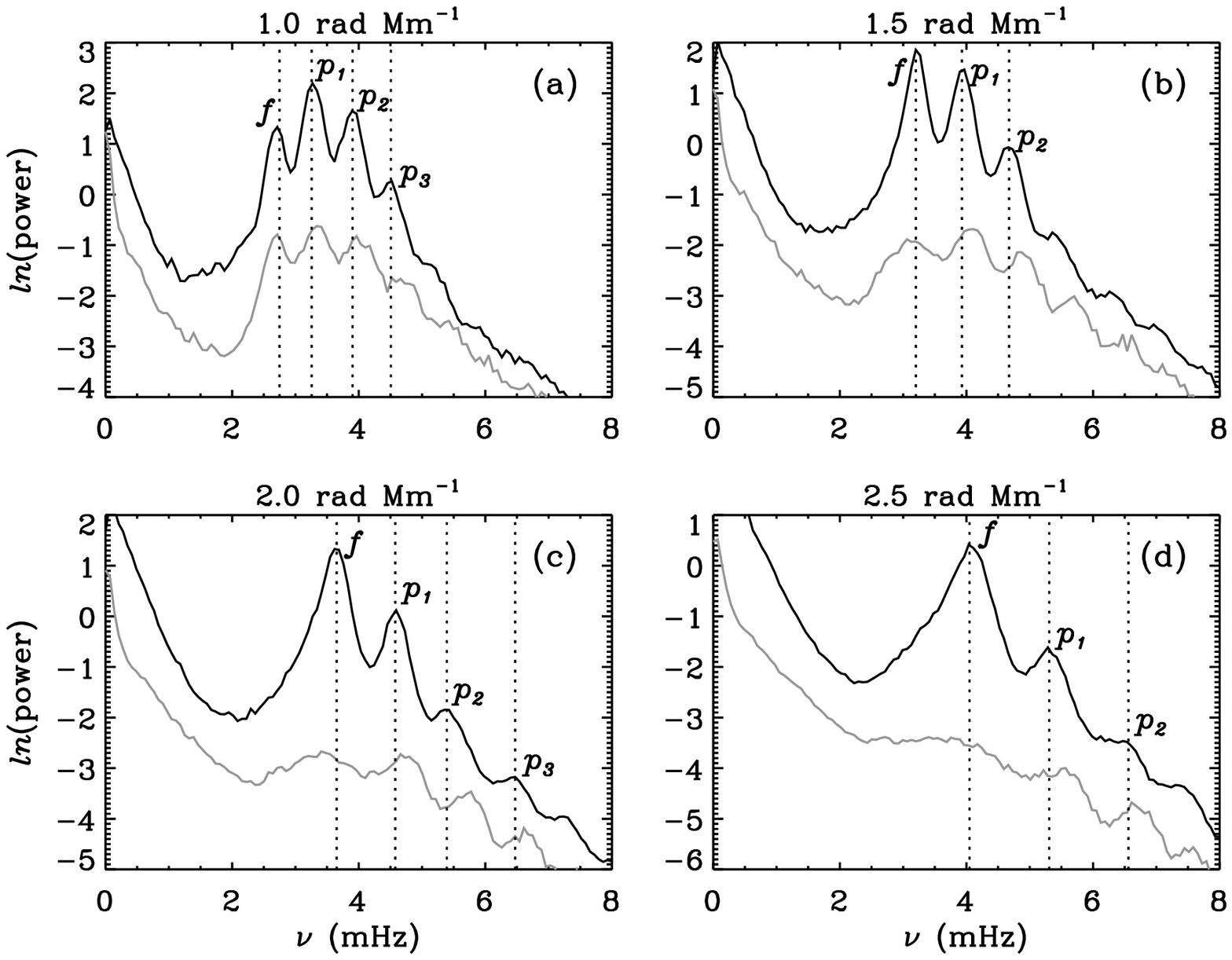} }
\caption{Comparisons of power spectrum at different wavenumbers obtained 
from the sunspot (gray lines) and the quiet-Sun region (black lines). 
The vertical dashed lines indicate the locations of the ridge peaks for 
the quiet Sun. } 
\label{kw_vcut}
\end{figure}

Comparisons of the ridge shape and location at selected frequencies are 
displayed in Figure~\ref{kw_cut}, where the curves are obtained by averaging 
a 0.2-mHz wide band in the \kw diagrams of Figure~\ref{kw}a and b.
Figure~\ref{kw_cut} again shows that the helioseismic power of the $f$-mode 
ridge is more suppressed in the sunspot. All of the $p$-mode ridges show 
clear shifts toward lower wavenumber, and the shifts are more significant 
in higher frequencies. The amount of ridge shift can be as large as half 
of the ridge-width for $p_1$, $p_2$, and $p_3$ ridges at the frequency of 
5.0 and 6.0~mHz, as can be seen in Figure~\ref{kw_cut}c and d. It can also 
be found in Figure~\ref{kw_cut} that the modal ridge line-profile asymmetry 
is more significant in the power spectrum from the sunspot than from the 
quiet-Sun region.  It is well known that the modal ridges obtained from 
Doppler observations exhibit line-profile asymmetries \cite{duv93}, and 
this asymmetry is believed to be caused by correlated noise \cite{nig98, 
geo03}. It is not immediately clear whether the more prominent line-profile 
asymmetries in sunspots can be explained by more substantial correlated
noise observed in these areas. On the other hand, recent studies 
(\eg\ \opencite{cho09}) found that the power absorption and local 
suppression of acoustic waves inside sunspots increase with wavenumber,
and the more prominent line-profile asymmetry may possibly be due to 
this effect.

Figure~\ref{kw_vcut} shows comparisons of ridge shapes and locations at
some selected wavenumbers. For each given wavenumber, the $p_1$--$p_3$
ridges all display shifts toward higher frequency, and the amount
of shift can also be up to half of the ridge width. The significant 
line-profile asymmetry is also clear in the $p$-mode ridges obtained from 
the sunspot region. For the $f$-mode ridge, it seems that the ridge 
shifts toward lower frequency in panels (c) and (d) of Figure~\ref{kw_vcut}, 
but that may just be due to the enhancement of background noises near 
these frequencies.

\begin{figure}[t]
\centerline{\includegraphics[width=0.75\textwidth]{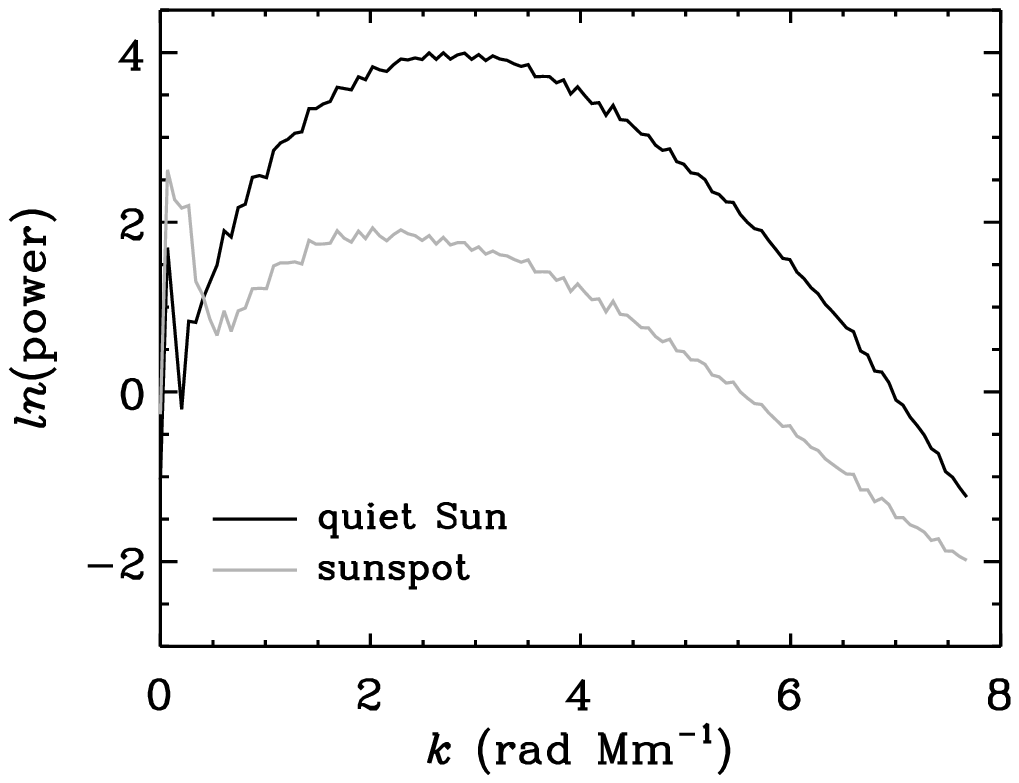} }
\caption{Comparison of convection power of the sunspot and the quiet-Sun
region. }
\label{kw_conv}
\end{figure}

It is also interesting to compare convection power of the sunspot and 
the quiet-Sun region. The convection powers, primarily below 1.5~mHz,
can be seen in the lower part of panels (a) and (b) of Figure~\ref{kw}.
Following procedures employed by \inlinecite{cho92} and \inlinecite{geo07},
we calculate convection power as a function of wavenumber by integrating 
power below the straight line separating oscillation power and convection
power, {\it i.e.} $\nu = 1.03 k$, where $\nu$ is in unit of mHz and $k$ in unit
of rad Mm$^{-1}$. Figure~\ref{kw_conv} shows the comparison of convection
powers obtained from both \kw diagrams. Not surprisingly, convection is 
greatly suppressed inside the sunspot. The location of the convection peak 
is also shifted toward lower wavenumber in the power spectrum obtained 
from the sunspot relative to that from the quiet Sun. For the quiet Sun, 
convection peaks at wavenumber of approximately 2.7~rad~Mm$^{-1}$ that 
is corresponding to a characteristic spatial scale of 2.3~Mm, presumably 
representing the scale of solar granules. For the sunspot, convection 
peaks at approximately 1.9~rad~Mm$^{-1}$, corresponding to a characteristic 
spatial scale of 3.3~Mm, over 40\% larger than the typical convection scale 
estimated for the quiet Sun. This is consistent with the result of 
time -- distance analysis of a sunspot observed by {\it Hinode} Ca~H line 
\cite{zha10} in that the convection scale is larger inside sunspots than in 
quiet Sun, but different in the convection cell sizes inside the sunspot, 
which was reported as 4 -- 5 Mm in that study. The difference in
sizes obtained from these two studies may be due to the different spectral 
line formation heights. The reason of why the 
convection cell size appears larger inside the sunspot is not clear, but 
it is possible that magnetic field suppresses convection more strongly 
in smaller scales and leaves larger-scale convection more prominent. 
And also, we cannot completely rule out the possibility of that the penumbral
filament, which is typically larger than granules, leaks into our analysis 
of convection power inside the sunspot. 

\section{Discussion}
\label{sec4}

\subsection{Possible Measurement Uncertainties}
We have constructed the \kw power-spectrum diagram using only oscillations
inside a sunspot observed by SDO/HMI. Distinct $f$- and 
$p$-mode ridges are clearly seen, and this demonstrates that resonant 
oscillations exist inside the sunspot, primarily inside the sunspot penumbra.
It is not clear how much scattered light, the light scattered into the dark
sunspot umbra from the brighter surrounding areas, plays a role in our 
computed \kw diagram. It is estimated that in an HMI-observed sunspot umbra,
the scattered light composes approximately $20\%$ of the light intensity
(T. Duvall, private communication, 2012). However, the power spectrum seen in
Figure~\ref{kw}a is mainly from the sunspot penumbra, and this is
clear from two experiments that we have performed. One is to mask out all 
oscillations in the umbra, and the other is to enhance the oscillations 
in the umbra to the same amplitude as the oscillations in the penumbra. 
In both experiments, the \kw diagram is not much different from the one 
shown in Figure~\ref{kw}a except in very low-$k$ area, where results
are already deemed not useful due to the selection of a small area 
and zero-padding. Therefore, the effect of scattered light to the computed 
\kw diagram of the sunspot is believed negligible in the diagram area 
that we are interested in. We also attempted to construct a \kw diagram 
by using only oscillations signals inside the sunspot umbra, but were
not able to detect distinct modal ridges, probably due to the small 
size of the umbra and its weak oscillation amplitude.

The uncertainties in determining Doppler velocities inside the sunspot
may also have some effects in the \kw\ diagram constructed in 
this study. \inlinecite{nor06} discussed the sensitivity of the line 
Fe~I~6173\AA\ to magnetic field, and \inlinecite{cou12} examined the 
uncertainties of deriving Doppler velocities, particularly inside active
regions, using different methods. The uncertainties in Doppler velocities
may cause some uncertainties in power measurements but are not expected
to play a significant role in shifting the locations of modal ridges
found in this study.  

\subsection{Implications for Sunspot Subsurface Structure}
It is interesting to understand why the $p$-mode ridges shift toward 
lower wavenumber (or higher frequency). \inlinecite{sch08} explored the 
wave motions with the presence of inclined magnetic field inside sunspot
penumbra, and \inlinecite{fel10} discussed the wave mode conversions in
the magnetized plasma by using 3D MHD simulations. More recently, 
\inlinecite{kit11} showed through MHD simulations that vertical 
magnetic field would introduce some acoustic frequency shifts toward 
higher frequency, but an inclined magnetic field would shift the frequency 
more substantially. This is consistent with our observation 
of modal-ridge shifts toward higher frequency in the sunspot, where 
the sunspot penumbra, the region with strong inclined magnetic field, 
contributes most to our computed \kw diagram. It is possible that the 
modal-ridge shift is due to that part of the acoustic waves converts
to fast magnetoacoustic waves when passing the magnetic field beneath
the sunspot, and the phase velocity of fast magnetoacoustic wave is
determined by:
\begin{equation} 
v_{\mathrm{ph}}^2=\frac{1}{2}(c_\mathrm{s}^2 
+ v_\mathrm{A}^2) + \frac{1}{2}\sqrt{(c_\mathrm{s}^2 + v_\mathrm{A}^2)^2 -
4 c_\mathrm{s}^2 v_\mathrm{A}^2 \cos^2\psi},
\label{eq1}
\end{equation} 
where $c_\mathrm{s}$, $v_\mathrm{A}$, and $\psi$ are sound speed, 
Alfv\'{e}n speed, and angle between directions of wave-vector and magnetic 
field, respectively.  The observed 
power spectrum is an aggregate from different depths beneath the sunspot,
so in principle, it is possible to invert the observed power spectrum
for the interior magnetic-field distribution although this inversion
procedure is considerably complicated. 

The relative change in phase velocity $\delta v_{\mathrm{ph}}/v_{\mathrm{ph}}$
shown in Figure~\ref{phv}b may provide some information beneath the sunspot.
The small values of $\delta v_{\mathrm{ph}}/v_{\mathrm{ph}}$,
which falls in a range of $0.02 - 0.06$, together with Equation~(\ref{eq1})
suggests that the averaged $\delta c_s/c_s$ and $v^2_A/c^2_s$ inside 
the sunspot is small, where $\delta c_s$ is the sound-speed change 
inside the sunspot relative to the quiet Sun. If $\delta c_s/c_s \ll 1$ and 
$v^2_A/c^2_s\ll 1$, from Equation~(\ref{eq1}), we have
\begin{equation}
\frac{\delta v_{\mathrm{ph}}}{v_{\mathrm{ph}}} \approx \frac{\delta c_s}
{c_s} + \frac{1}{2} \sin^2\psi \left(\frac{v^2_A}{c^2_s}\right).
\label{eq2}
\end{equation}
The observed value of $\delta v_{\mathrm{ph}}/v_{\mathrm{ph}}$ gives a 
constraint on the value of averaged $\delta c_s/c_s + \sin^2\psi 
(v^2_A/c^2_s)$, and that is $0.02 < \delta c_s/c_s + \sin^2\psi (v^2_A/c^2_s)
< 0.06$. It is noted that $\delta c_s/c_s + \sin^2\psi 
(v^2_A/c^2_s)$ is averaged over the three-dimensional sunspot region 
weighted with a kernel, which is proportional to the mode energy density
though the mode energy density inside the sunspot is unknown. 
Unlike $\sin^2\psi (v^2_A/c^2_s)$ which is always positive, $\delta c_s/c_s$
could be positive or negative beneath the sunspot. Since 
$\delta c_s/c_s \sim \delta\gamma/2\gamma + \delta T/2T$ and $v_A^2/2c_s^2
=1 /\beta$, where $\gamma$ is the adiabatic index and $\beta$ the ratio
of gas pressure to magnetic pressure. $|\delta c_s/c_s|$ is smaller 
than $\sin^2\psi (v^2_A/c^2_s)$ near the surface if $\sin^2\psi$ is not
too small. Thus, most of contribution to $\delta v_\mathrm{ph}/v_\mathrm{ph}$
could be from $\sin^2\psi (v^2_A/c^2_s)$. Another possible contribution
to $\delta v_\mathrm{ph}/v_\mathrm{ph}$, which is not shown in 
Equation~(\ref{eq1}),
is the change in the mode cavity, although it could be smaller than that 
in Equation~(\ref{eq2}). The upper turning point inside the sunspot is deeper 
than in the quiet Sun, thus its contribution to $\delta v_\mathrm{ph}
/v_\mathrm{ph}$ is positive.

An interesting phenomenon shown in Figure~\ref{phv}b is that
$\delta v_{\mathrm{ph}}/v_{\mathrm{ph}}$ has a general trend for each $n$: it
first increases and then decreases with $k$. This could 
be explained by Equation~(\ref{eq2}). It is generally believed that $v^2_A/c^2_s$ 
is significant only near the surface. Since the mode of higher $k$ 
penetrates less deeply into the solar interior, averaged $v^2_A/c^2_s$ 
increases with $k$. On the other hand, the mode of higher $k$ propagates 
less vertically near the surface, and the magnetic field inside the sunspot 
penumbra has a large horizontal component near the surface. This leads to 
that a decrease of $\psi$ with $k$, or averaged $\sin^2\psi$ increases with $k$.
Therefore, averaged $v^2_A/c^2_s$ increases with $k$ while averaged 
$\sin^2\psi$ decreases with $k$. This yields a maximum of $\sin^2\psi 
(v^2_A/c^2_s)$ or $\delta v_{\mathrm{ph}}/v_{\mathrm{ph}}$ at some $k$.

\subsection{Relevance to Earlier Studies}
Recent power-spectral analysis by \inlinecite{sch11} showed that the modal
ridges obtained from the sunspot vicinity, where acoustic halos are 
detectable, shift to higher-wavenumber areas relative to the quiet Sun. 
That shift is in a direction opposite to what we find inside the sunspot 
area in this study. It seems that the modal ridges shift toward lower 
wavenumber in areas where acoustic power gets reduced, and shift toward 
higher wavenumber in areas where acoustic power gets enhanced. Moreover, 
the phase velocity becomes faster in the sunspot penumbra possibly 
implying the presence of faster magnetoacoustic waves, and phase velocity 
becomes slower in the sunspot vicinity possibly implying a presence 
of slower magnetoacoustic waves in the area. It is interesting to note
that in that same study by \inlinecite{sch11}, the modal-ridge lines 
showed less line-asymmetry than in the quiet-Sun region, again in an 
opposite sense to what is found inside a sunspot in this study. This
is also a curious phenomenon, but it is not clear whether these opposite
line-asymmetry trends are related to the different atmospheric heights
of sunspot penumbra and sunspot vicinity, or related to the faster
and slower magnetoacoustic waves discussed above. 

The modal-ridge shift observed inside the sunspot may complicate
the interpretation of the measured travel-time shifts or phase shifts 
in sunspot areas with an application of phase-velocity filter or a 
modal-ridge filter. Because these filters choose parameters based on 
quiet-Sun areas, the power ridges of sunspots may not be filtered in
the same way as the power ridges of quiet-Sun areas due to that the 
modal ridges of sunspots are systematically located on lower-wavenumber
(or higher-frequency) side. However, we believe the effect may not be 
significant,
as \inlinecite{zha10} showed that the measurements with and without using 
phase-speed filters gave results in a good qualitative agreement. This 
observed modal-ridge shift in sunspot regions may not have an effect
in the ring-diagram derived subsurface properties in active regions, such
as given by \inlinecite{bal11}, because acoustic power of the sunspot
contributed little to the power spectrum used to do ring-diagram 
analysis that usually analyzes a much larger area than a sunspot.

\section{Conclusion}
\label{sec5}

By use of 40 hours of continuous SDO/HMI Doppler observations of a stable 
sunspot, we construct a \kw power-spectrum diagram that shows distinct 
modal ridges with suppressed 
helioseismic power. More notably, comparing with the \kw diagram obtained 
from a quiet-Sun region, the $f$-mode ridge in the diagram of the sunspot
gets more suppressed in power than $p$-mode ridges, especially at
high wavenumber. All $p$-mode ridges shift toward lower-wavenumber
(or higher-frequency) area for a given frequency (or wavenumber). The phase
velocity computed from our fitted $p$-mode ridges shows an increase of 
$2-6\%$ inside the sunspot region. Inclined magnetic field in the sunspot 
penumbra may be responsible for the $p$-mode ridge shifts
and the increase of the phase velocity. The line-profile asymmetries in 
the $p$-mode ridges for the sunspot region are more prominent than the 
asymmetries observed in quiet Sun, and this may be relevant to the higher
power absorption and suppression ratio in higher wavenumbers inside 
the sunspot. Convection inside the sunspot is also highly suppressed, 
but exhibits a characteristic spatial scale approximately $40\%$ larger than 
the typical granulation scale in the quiet Sun. Realistic MHD simulations 
of magnetoconvection with helioseismic waves included (\eg\ \opencite{rem09}) 
are very useful to understand all of these observed facts inside the 
sunspot region, and we expect that an analysis similar to what is presented
in this study but using numerical simulation data will shed new
light in a better understanding of the sunspot subsurface structure. 

\begin{acks}
SDO is a NASA mission, and HMI project is supported by NASA contract
NAS5-02139. We thank Mark Cheung for useful discussions on convection
scales inside sunspots. We also thank the anonymous referee for useful
comments to improve the quality of this paper.
\end{acks}

\end{article}

\begin{thebibliography}

\bibitem[\protect\citeauthoryear{{Baldner}, {Bogart}, and {Basu}}{2011}]{bal11}
Baldner, C.S., Bogart, R.S., Basu, S.: 2011, {\it J. Phys. Conf. Ser.} 
\textbf{271}, 012006.

\bibitem[\protect\citeauthoryear{{Basu}, {Antia}, and {Bogart}}{2004}]{bas04}
Basu, S., Antia, H.M., Bogart, R.S.: 2004, \apj{} \textbf{610}, 1157.

\bibitem[\protect\citeauthoryear{{Braun}, {Duvall}, and {LaBonte}}{1987}]{bra87}
Braun, D.C., Duvall, T.L. Jr., LaBonte, B.J.: 1987, \apj{} \textbf{319}, L27.

\bibitem[\protect\citeauthoryear{Chou \etal}{1992}]{cho92} Chou, D.-Y.,
Chen, C.-S., Ou, K.-T., Wang, C.-C.: 1992, \apj{} \textbf{396}, 333.

\bibitem[\protect\citeauthoryear{Chou \etal}{2009}]{cho09} Chou, D.-Y.,
Yang, M.-H., Zhao, H., Liang, Z.-C., Sun, M.-T.: 2009, \apj{} \textbf{706}, 909.

\bibitem[\protect\citeauthoryear{Couvidat \etal}{2012}]{cou12} Couvidat, S.,
Rajaguru, S.P., Wachter, R., Sankarasubramanian, K., Schou, J., Scherrer, P.H.:
2012, \solphys{} \textbf{278}, 217. ADS:\adsurl{2012SoPh..278..217C},
doi:\doiurl{10.1007/s11207-011-9927-y}

\bibitem[\protect\citeauthoryear{Duvall \etal}{1993a}]{duv93} Duvall, T.L. Jr.,
Jefferies, S.M., Harvey, J.W., Osaki, Y., Pomerantz, M. A.: 1993a, \apj{}
\textbf{410}, 829.

\bibitem[\protect\citeauthoryear{Duvall \etal}{1993b}]{duv93b} Duvall, T.L. Jr.,
Jefferies, S.M., Harvey, J.W., Pomerantz, M. A.: 1993b, \nat{} \textbf{362},
430.

\bibitem[\protect\citeauthoryear{{Felipe}, {Khomenko}, and {Collados}}{2010}]
{fel10} Felipe, T., Khomenko, E., Collados, M.: 2010, \apj{} \textbf{719}, 357.

\bibitem[\protect\citeauthoryear{{Georgobiani}, {Stein}, and {Nordlund}}
{2003}]{geo03} Georgobiani, D., Stein, R.F., Nordlund, \AA.: 2003, \apj{}
\textbf{596}, 698.

\bibitem[\protect\citeauthoryear{Georgobiani \etal}{2007}]{geo07} Georgobiani, 
D., Zhao, J., Kosovichev, A.G., Benson, D., Stein, R.F., Nordlund, \AA.: 
2007, \apj{} \textbf{657}, 1157.

\bibitem[\protect\citeauthoryear{Gizon \etal}{2009}]{giz09} Gizon, L.,
Schunker, H., Baldner, C.S., Basu, S., Birch, A.C., Bogart, R.S., Braun, D.C.,
Cameron, R., Duvall, T.L., Hanasoge, S.M., \etal: 2009, \ssr{} \textbf{144}, 
249.

\bibitem[\protect\citeauthoryear{Haber \etal}{2004}]{hab04} Haber, D.A., 
Hindman, B.W., Toomre, J., Thompson, M.J.: 2004, \solphys{}, \textbf{220},
371. ADS:\adsurl{2004SoPh..220..371H},
doi:\doiurl{10.1023/B:SOLA-0000031405-52911-08}.

\bibitem[\protect\citeauthoryear{Hill}{1998}]{hil88} Hill, F.: 1988,
\apj{} \textbf{333}, 996.

\bibitem[\protect\citeauthoryear{Howard}{1990}]{how90} Howard, R.F.:
1990, \solphys{}, \textbf{126}, 299. ADS:\adsurl{1990SoPh..126..299H},
doi:\doiurl{10.1007/BF00153052}.


\bibitem[\protect\citeauthoryear{Kitiashvili \etal}{2011}]{kit11}
Kitiashvili, I.N., Kosovichev, A.G., Mansour, N.N., Wray, A.A.: 2011,
\solphys{}, \textbf{268}, 283. ADS:\adsurl{2011SoPh..268..283K},
doi:\doiurl{10.1007/s11207-010-9679-0}.

\bibitem[\protect\citeauthoryear{Komm \etal}{2005}]{kom05} Komm, R., Howe, R.,
Hill, F., Gonz\'{a}lez Hern\'{a}ndez, I., Toner, C., Corbard, T.: 2005, 
\apj{} \textbf{631}, 636.

\bibitem[\protect\citeauthoryear{{Kosovichev}, {Duvall}, and {Scherrer}}{2000}]
{kos00} Kosovichev, A.G., Duvall, T.L. Jr., Scherrer, P.H.: 2000, \solphys{}
\textbf{192}, 159. ADS:\adsurl{2000SoPh..192..159K},
doi:\doiurl{10.1023/A:1005251208431}.

\bibitem[\protect\citeauthoryear{Nagashima \etal}{2007}]{nag07} Nagashima, K.,
Sekii, T., Kosovichev, A.G., Shibahashi, H., Tsuneta, S., Ichimoto, K.,
Katsukawa, Y., Lites, B., Nagata, S., Shimizu, T., \etal: 2007, \pasj{}
\textbf{59}, S631.

\bibitem[\protect\citeauthoryear{Nigam \etal}{1998}]{nig98} Nigam, R.,
Kosovichev, A.G., Scherrer, P.H., Schou, J.: 1998, \apj{} \textbf{495}, L115.

\bibitem[\protect\citeauthoryear{Norton \etal}{2006}]{nor06} Norton, A.A.,
Graham, J.P., Ulrich, R.K., Schou, J., Tomczyk, S., Liu, Y., Lites, B.W.,
L\'{o}pez Ariste, A., Bush, R.I., Socas-Navarro, H., Scherrer, P.H.: 
2006, \solphys{} \textbf{239}, 69. ADS:\adsurl{2006SoPh..239...69N},
doi:\doiurl{10.1007/s11207-006-0279-y}.

\bibitem[\protect\citeauthoryear{{Penn} and {LaBonte}}{1993}]{pen93} Penn, M.J.,
LaBonte, B.J.: 1993, \apj{} \textbf{415}, 383.

\bibitem[\protect\citeauthoryear{Rempel \etal}{2009}]{rem09} Rempel, M.,
Sch\"{u}ssler, M., Cameron, R.H., Kn\"{o}lker, M.: 2009, {\it Science} 
\textbf{325}, 171.

\bibitem[\protect\citeauthoryear{Rhodes \etal}{1997}]{rho97} Rhodes, E.J., Jr.,
Kosovichev, A.G., Schou, J., Scherrer, P.H., Reiter, J.: 1997, \solphys{}
\textbf{175}, 287. ADS:\adsurl{1997SoPh..175..287R},
doi:\doiurl{10.1023/A:1004963425123}.

\bibitem[\protect\citeauthoryear{Scherrer \etal}{1995}]{sch95} Scherrer, P.H.,
Bogart, R.S., Bush, R.I., Hoeksema, J.T., Kosovichev, A.G., Schou, J.,
Rosenberg, W., Springer, L.; Tarbell, T.D., Title, A., \etal: 1995, 
\solphys{} \textbf{162}, 129. ADS:\adsurl{1995SoPh..162..129S},
doi:\doiurl{10.1007/BF00733429}.

\bibitem[\protect\citeauthoryear{Scherrer \etal}{2012}]{scherrer12} 
Scherrer, P.H., Schou, J., Bush, R.I., Kosovichev, A.G., Bogart, R.S., 
Hoeksema, J.T., Liu, Y., Duvall, T.L. Jr., Zhao, J., Title, A.M., \etal: 
2012, \solphys{} \textbf{275}, 207. ADS:\adsurl{2012SoPh..275..207S},
doi:\doiurl{10.1007/s11207-011-9834-2}

\bibitem[\protect\citeauthoryear{Schou \etal}{1998}]{sch98} Schou, J.,
Antia, H.M., Basu, S., Bogart, R.S., Bush, R.I., Chitre, S.M., 
Christensen-Dalsgaard, J., di Mauro, M.P., Dziembowski, W.A., Eff-Darwich, A.,
\etal: 1998, \apj{} \textbf{505}, 390.

\bibitem[\protect\citeauthoryear{Schou \etal}{2012}]{sch12} Schou, J., 
Scherrer, P.H., Bush, R.I., Wachter, R., Couvidat, S., Rabello-Soares, M.C.,
Bogart, R.S., Hoeksema, J.T., Liu, Y., Duvall, T.L. Jr., \etal: 2012,
\solphys{} \textbf{275}, 229. ADS:\adsurl{2012SoPh..275..229S},
doi:\doiurl{10.1007/s11207-011-9842-2}.

\bibitem[\protect\citeauthoryear{{Schunker} and {Braun}}{2011}]{sch11} 
Schunker, H., Braun, D.C.: 2011, \solphys{} \textbf{268}, 349.
ADS:\adsurl{2011SoPh..268..349S},
doi:\doiurl{10.1007/s11207-010-9550-3}.

\bibitem[\protect\citeauthoryear{Schunker \etal}{2008}]{sch08} Schunker, H.,
Braun, D.C., Lindsey, C., Cally, P.S.: 2008, \solphys{} \textbf{251}, 341.
ADS:\adsurl{2008SoPh..251..341S},
doi:\doiurl{10.1007/s11207-008-9142-7}.

\bibitem[\protect\citeauthoryear{Zhao \etal}{2012}]{zha12} Zhao, J., 
Couvidat, S., Bogart, R.S., Parchevsky, K.V., Birch, A.C., Duvall, T.L. Jr.,
Beck, J.G., Kosovichev, A.G., Scherrer, P.H.: 2012, \solphys{} \textbf{275},
375. ADS:\adsurl{2012SoPh..275..375Z},
doi:\doiurl{10.1007/s11207-011-9757-y}.

\bibitem[\protect\citeauthoryear{{Zhao}, {Kosovichev}, and {Sekii}}{2010}]
{zha10} Zhao, J., Kosovichev, A.G., Sekii, T.: 2010, \apj{} \textbf{708}, 304.

\end{thebibliography}
\end{document}